\documentclass[11pt]{article}
\usepackage{moriond,epsfig}

\RequirePackage{xspace}
\newcommand{ \nOne} {\mbox{$\tilde{\chi}_{1}^0$}}
\newcommand{ \cOne }{\mbox{$\tilde{\chi}_{1}^{\pm}$}}
\newcommand{ \nTwo }{\mbox{$\tilde{\chi}_{2}^0$}}

\newcommand{ \VLQ}  {\mbox{$V\!LQ3$}}
\newcommand{ \SLQthree}  {\mbox{$S\!LQ3$}}
\newcommand{ \SLQtwo}  {\mbox{$S\!LQ2$}}

\newcommand{\met}{\mbox{${E\!\!\!\!/_T}$}}
\def\pt         {\mbox{$p_T$}\xspace}
\def\Et         {\mbox{$E_T$}\xspace}
\newcommand{\gev}{\ensuremath{\mathrm{\,Ge\kern -0.1em V}}\xspace}
\newcommand{\gevc}{\ensuremath{{\mathrm{\,Ge\kern -0.1em V\!/}c}}\xspace}
\newcommand{\gevcc}{\ensuremath{{\mathrm{\,Ge\kern -0.1em V\!/}c^2}}\xspace}
\newcommand{\tev}{\ensuremath{\mathrm{\,Te\kern -0.1em V}}\xspace}
\def\invpb {\ensuremath{\mbox{\,pb}^{-1}}\xspace}
\def\invfb   {\ensuremath{\mbox{\,fb}^{-1}}\xspace}
\def\ns         {\ensuremath{{\rm \,ns}}\xspace}      %% nanosecond
\def\mm   {\ensuremath{\rm \,mm}\xspace}

\def\be{\begin{equation}}
\def\ee{\end{equation}}
\def\bea{\begin{eqnarray}}
\def\eea{\end{eqnarray}}

\begin{document}
\vspace*{4cm}
\title{SEARCHES IN PHOTON AND JET STATES}

\author{ A.~SOHA \\
FOR THE CDF AND D\O\ COLLABORATIONS}

\address{University of California, Department of Physics \\
One Shields Avenue, Davis, California 95616, USA}

\maketitle\abstracts{We present recent results from the Collider Detector at
Fermilab (CDF) and D\O\ experiments using data from proton-antiproton
collisions with $\sqrt{s} = 1.96 \tev$ at Run II of the Fermilab Tevatron.
New physics may appear in events with high transverse momentum objects,
including photons and quark or gluon jets.  The results described here
are of signature-based searches and model-based
searches probing supersymmetry, leptoquarks, $4^{\rm th}$ generation
quarks, and large extra dimensions.}

\section{Signature Based Searches}

\subsection{Searches in $\gamma\gamma$ + $\{e$, $\mu$, $\gamma$, $\met\}$}

The signature-based search for two photons plus an additional
photon, electron, muon, or missing transverse energy ($\met$) at CDF
in Run II~\cite{ggX} is motivated by an anomalous Run I event~\cite{anom}
containing $ee\gamma\gamma\met$, and the fact that new physics may appear
in events with multiple high transverse momentum ($\pt$) particles.
This signature may occur in numerous scenarios beyond the standard
model (SM):  pair-production of supersymmetry (SUSY) particles that yield
multiple photons as
they decay to lower mass states (for example, through $\nTwo\to\gamma\nOne$);
pair-production and radiative decays of other excited states
($X^{*}X^{*}\to\gamma X \gamma X$); $b'$ quark pair production and decay;
fermiphobic Higgs decays yielding photons; and gauge-mediated supersymmetry
breaking (GMSB) scenarios.  Using data corresponding to an integrated
luminosity of $1.0-1.2 \invfb$, the number of observed events
($4, 1, 3, 0$) in each of the four final states
($\gamma\gamma\gamma, \gamma\gamma\met, \gamma\gamma e, \gamma\gamma\mu$)
is consistent with the expected number of events due to SM
processes ($2.2, 0.24, 6.8, 0.7$).

\subsection{Searches in $\ell\gamma$ + $\{\ell$, $\gamma$, $\met\}$}

An additional signature-based analysis at CDF searches for events containing
an electron or muon ($\ell = e$ or $\mu$) with $\pt > 25 \gevc$,
a photon with $\Et > 25 \gev$, and an additional object, which can be
an electron, muon, photon, or $\met$~\cite{LepPhoX}.
This search, which uses $929 \invpb$, is motivated by a $2.7 \sigma$
excess found in Run I in the $\ell\gamma\met$ final state~\cite{RunILepPhoX}.  
In that search, $16$ events were observed compared to an expectation of
$7.6 \pm 0.7$ events due to SM sources.  The current search uses the
same event selection criteria, but about $10$ times the amount of data,
and observes $163$ $\ell\gamma\met$ events compared to a SM expectation
of $150.6 \pm 13.0$ events.  Table~\ref{table:LepPhoX} shows the
expected and observed number of events for the individual channels,
as well as the dominant SM process which yields each signature.
In all cases, the observations are consistent with the SM.

\begin{table}[t]
\caption{Number of expected and observed events for the
$\ell\gamma$ + $\{\ell$, $\gamma$, $\met\}$ search, and the dominate
SM processes.  \label{table:LepPhoX}}
\vspace{0.4cm}
\begin{center}
\begin{tabular}{|l|cc|c|cc|cc|}
\hline
 & $e e\gamma$ & $\mu\mu\gamma$ & $e\mu\gamma + X$ &
 $e\gamma\gamma$ & $\mu\gamma\gamma$ & $e\gamma\met$ & $\mu\gamma\met$ \\
\hline
Expected (SM) & $39 \pm 5$ & $26 \pm 3$ & $1.0 \pm 0.3$ &
 $0.5 \pm 0.1$ & $0.1 \pm 0.1$ & $95 \pm 8$ & $56 \pm 7$ \\
Dominant Source & $Z\gamma$ & $Z\gamma$ & $Z\gamma$ &
 $e$ fakes $\gamma$ & jet fakes $\gamma$ & $W\gamma$ & $W\gamma$ \\
Observed & $53$ & $21$ & $0$ & $0$ & $0$ & $96$ & $67$ \\
\hline
\end{tabular}
\end{center}
\end{table}

\section{Supersymmetry}

\subsection{Gauge Mediated Supersymmetry Breaking in $\gamma\gamma\met$}
\label{subsec:GMSB}

\begin{figure}
\psfig{figure=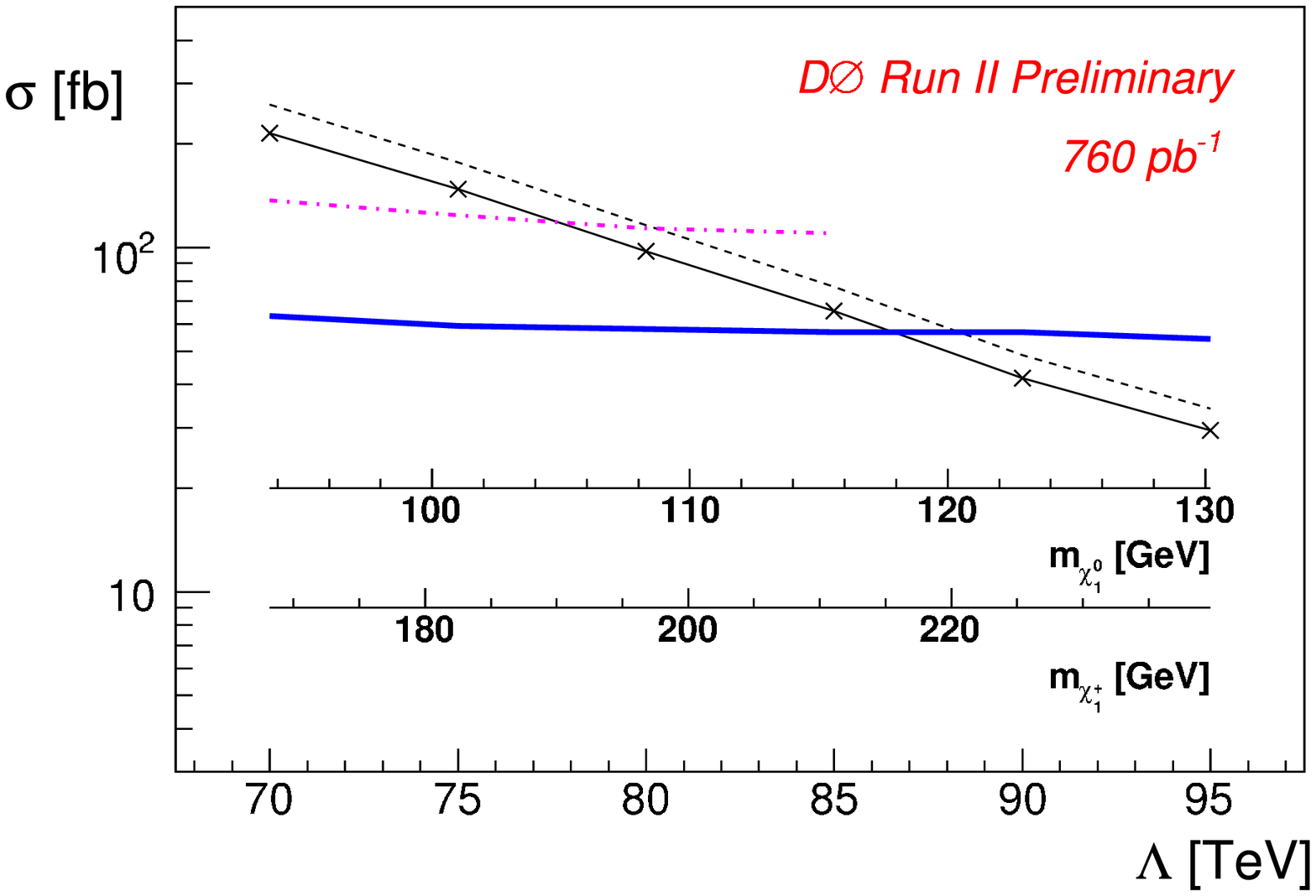,height=2.5in}
\psfig{figure=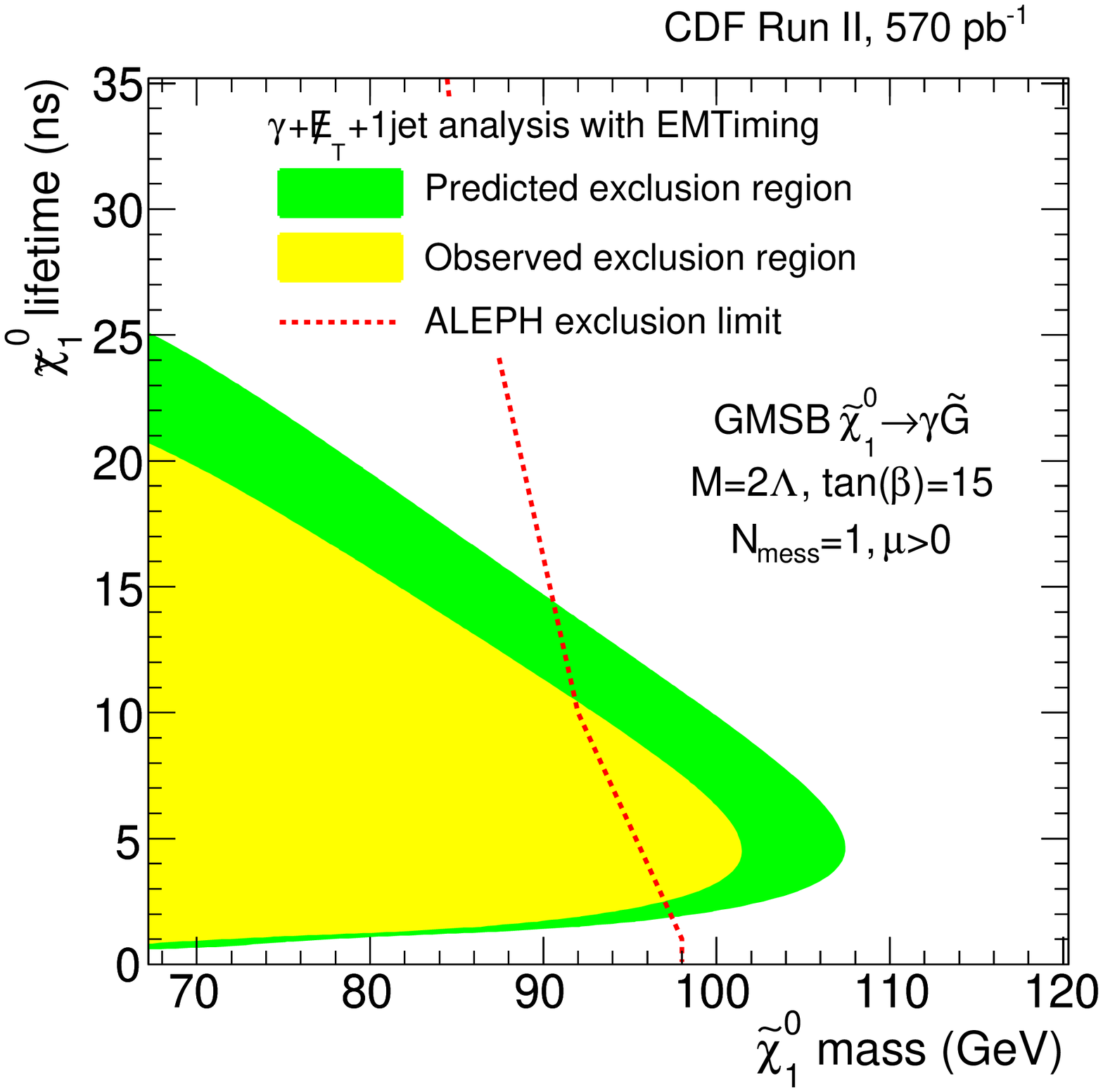,height=2.5in}
\caption{(left) Cross section limit for GMSB scenario as a function of SUSY
breaking scale, assuming prompt photons.  (right) Exclusion regions for a GMSB
scenario with delayed photons.
\label{fig:GMSBlimit}}
\end{figure}

% D0 note 5020-CONF
The spectrum of particles predicted by SUSY theory, which is a popular SM
extension, depends upon the mechanism by which SUSY is broken.
One possibility is the GMSB scenario, where
$\nOne\nOne\to\gamma\tilde{G}\gamma\tilde{G}$, yielding a signature of two
photons and $\met$ from the gravitinos ($\tilde{G}$).  A search using
$760 \invpb$ at D\O\ expects
$2.1 \pm 0.7$ events from SM sources and observes $4$ events~\cite{GMSB}.
Assuming short neutralino lifetime, so that the photons are produced
promptly, $95\%$ C.L.\ limits are set at $m(\nOne) > 120 \gevcc$ and
$m(\cOne) > 220 \gevcc$.  Figure~\ref{fig:GMSBlimit}(left) shows the cross
section limit as a function of the SUSY breaking scale parameter $\Lambda$.
The other parameters of the model are fixed at the so-called SPS8
values~\cite{SPS8}.

\subsection{Search for Delayed Photons}

The detection of photons produced after a delay with respect to the primary
interaction may indicate the presence of new phenomena.  These photons may
stem from decays of heavy, neutral, long-lived particles, such as in the
GMSB scenario, with $\nOne\to\gamma\tilde{G}$.  Using $570 \invpb$, CDF
searches for delayed photons in the $\gamma\met$+jet
channel~\cite{delayedPhotons}.  The measurement
uses a new timing system in the electromagnetic calorimeter.  The signal
selection requires the photon time of arrival to be $2 < t < 10 \ns$, reducing
backgrounds from beam halo, cosmic rays, and prompt photons.
We expect $1.3 \pm 0.7$ SM events and observe $2$ events.
Figure~\ref{fig:GMSBlimit}(right) shows the resulting exclusion regions in the
plane of neutralino lifetime and mass, assuming SPS8 parameters as in
section~\ref{subsec:GMSB}.

\subsection{Search for Stopped Gluinos}

\begin{figure}
\psfig{figure=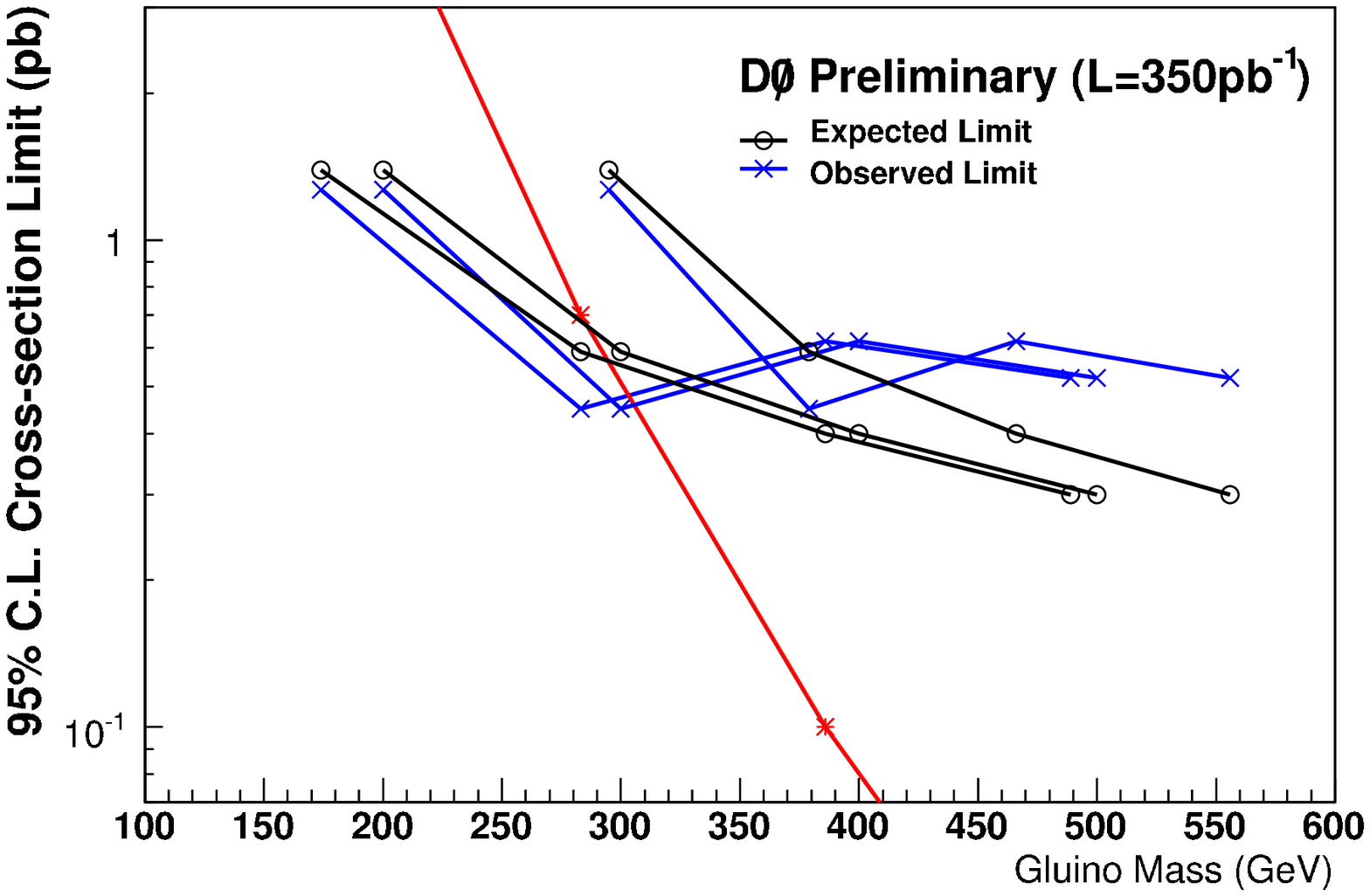,height=2.55in}
\psfig{figure=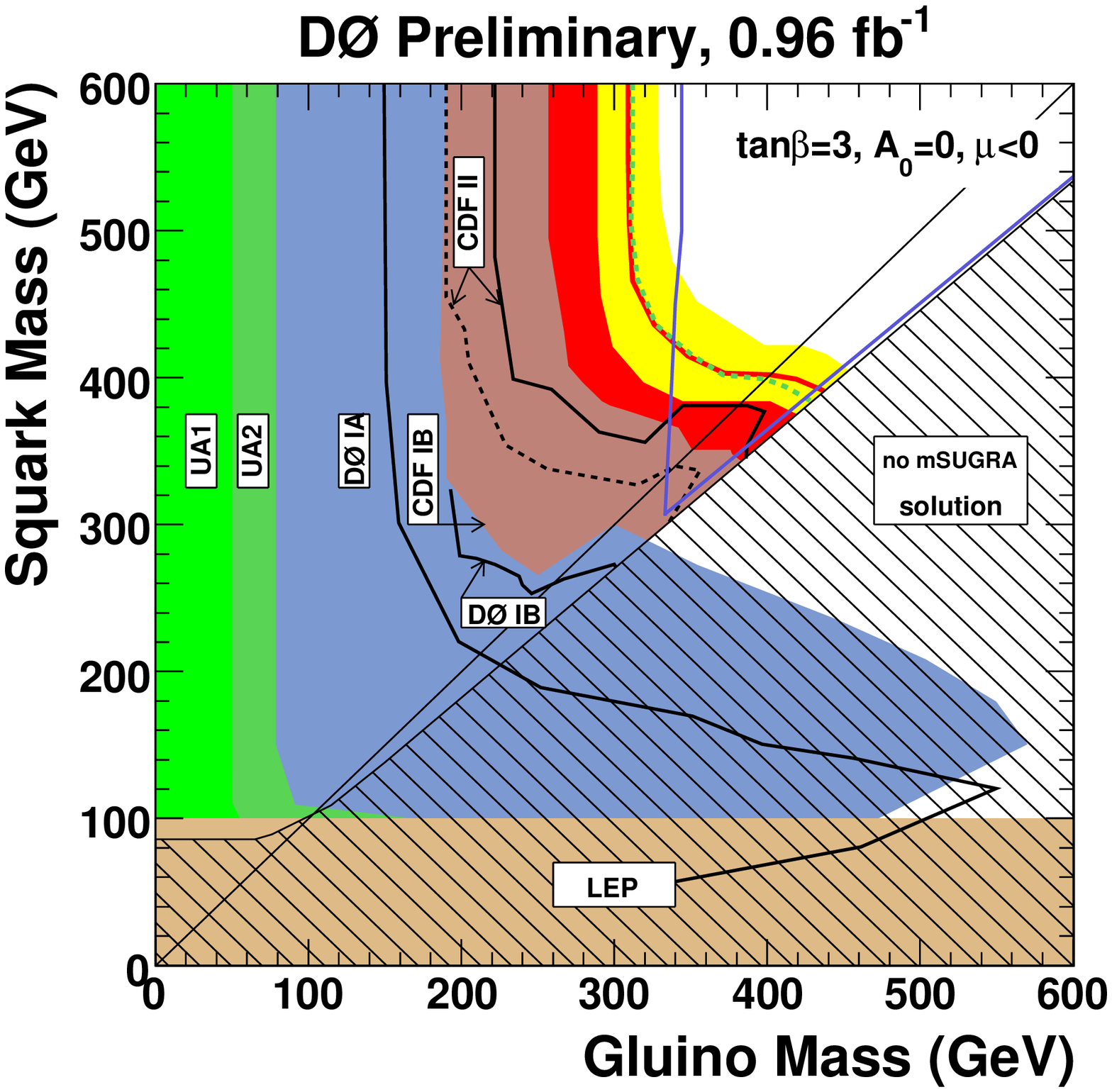,height=2.55in}
\caption{(left) Expected and observed gluino production cross section as a
function of gluino mass for stopped gluino scenarios with
$m(\nOne)=50,90,200 \gevcc$ (pairs of black and blue lines, left to right)
and the theoretical cross section (red line).
(right) Excluded regions for the squark and gluino search, showing
the newly excluded region in red and associated theoretical uncertainty
in yellow.
\label{fig:gluino}}
\end{figure}

In split-SUSY~\cite{splitSUSY}, the SUSY scalers are heavy (possibly of the
grand unification
scale) compared to the SUSY fermions.  Split-SUSY models predict long-lived
gluinos ($\tilde{g}$), which could have time to hadronize, lose momentum
through ionization, and come to rest in the calorimeter.  Using $350 \invpb$,
D\O\ searches for $\tilde{g}\to g\nOne$ in the jet+$\met$
channel~\cite{stoppedGluinos}, where the
gluon yields a jet and the unobserved $\nOne$ results in $\met$.
The selection requires one jet with $E > 90 \gev$, in an otherwise empty event.
Background from cosmic ray muons is reduced by requiring wide jet showers in
the calorimeter with no associated muon, since jets are wide for the
hypothetical signal but narrow for muons.  Additional beam-induced background
is successfully removed.  Cross section limits are set as a function of
jet energy and translated into gluino mass limits.  The search
sets a limit of $m(\tilde{g}) > \, \sim 270 \gevcc$ for an example with
$m(\nOne) = 50 \gevcc$.  Figure~\ref{fig:gluino}(left) shows cross section
limits as functions of gluino mass.

\subsection{Search for Squarks and Gluinos}

In the minimal supergravity (mSUGRA) formulation of SUSY, pair produced
squarks and gluinos
($\tilde{q}\tilde{q}, \tilde{q}\tilde{g}, \tilde{g}\tilde{g}$)
can decay through $\tilde{q}\to q\nOne$ and $\tilde{g}\to q\tilde{q}\nOne$
to yield signatures with 2, 3, or 4 jets from the quarks and
$\met$ from the neutralinos.  A search at D\O\ combines these three
channels~\cite{SqGl}, categorized by the number
of jets, to set $95\%$ C.L.\ limits on the squark and gluino masses
of $m(\tilde{q}) > 375 \gevcc$ and $m(\tilde{g}) > 290 \gevcc$,
using $960 \invpb$.  Figure~\ref{fig:gluino}(right) shows the excluded
regions in the squark--gluino mass plane.

\section{Leptoquarks}

\subsection{$3^{\rm rd}$ Generation Vector Leptoquarks}

\begin{figure}
\psfig{figure=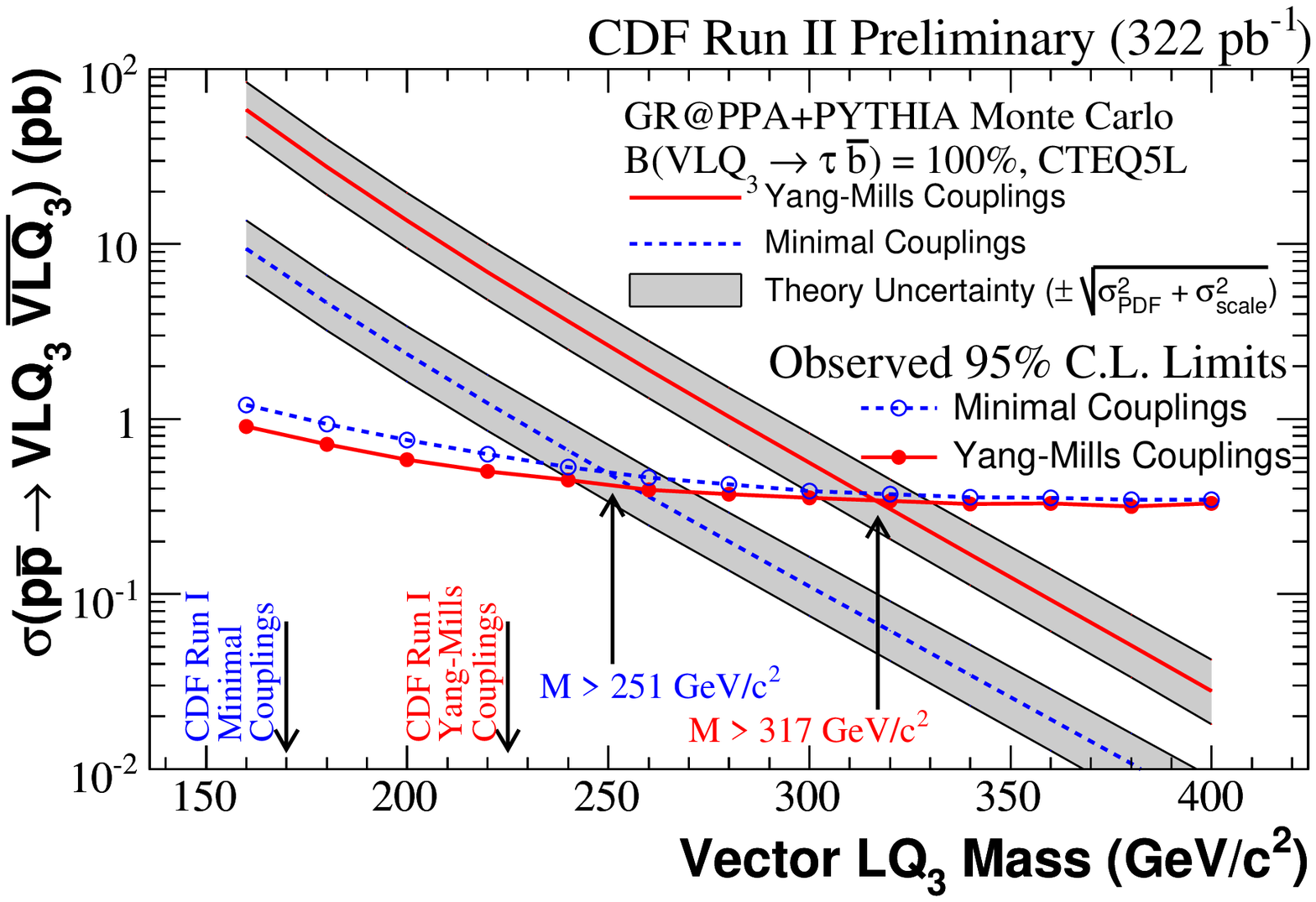,height=2.4in}
\psfig{figure=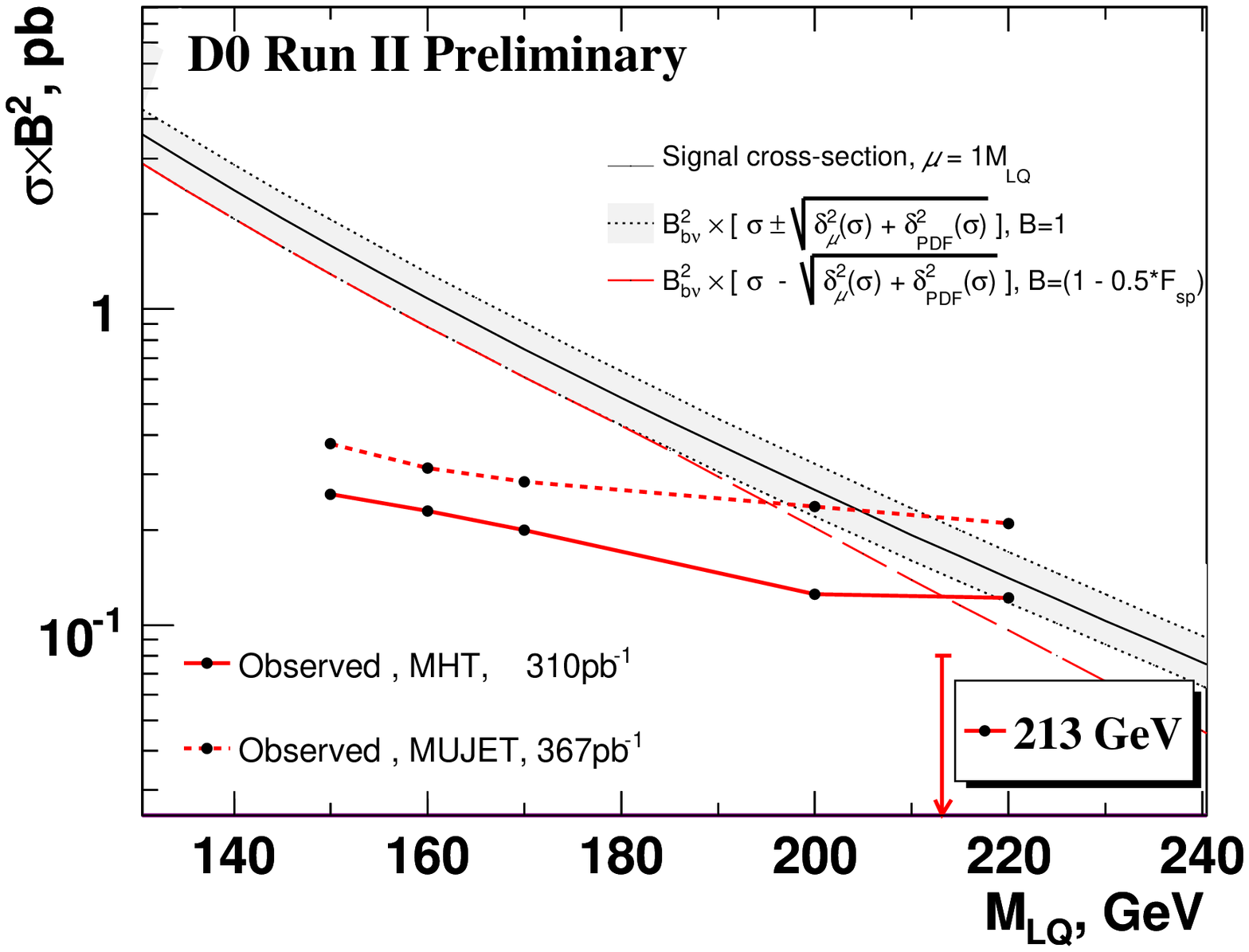,height=2.2in}
\caption{(left) Cross section versus mass for $3^{\rm rd}$ generation
vector leptoquarks, showing observed limits and theoretical predictions
for two coupling scenarios.  (right) Limits for $3^{\rm rd}$ generation
scalar leptoquarks as a function of $\SLQthree$ mass.  The results
quoted in the text are for the observed limit called MHT in this figure.
\label{fig:lq3_and_lq2}}
\end{figure}

Leptoquarks may provide a link between the families of quarks and leptons
at higher mass scales.  They appear in a variety of beyond the SM theories.
At the Fermilab Tevatron, leptoquarks could be predominately pair produced
through quark anti-quark annihilation.  Using $322 \invpb$ of CDF data, we
consider the case of $3^{\rm rd}$ generation vector leptoquarks (\VLQ),
where each decays exclusively to $\tau b$~\cite{lq3cdf}.  The decay products,
$\tau b \tau \overline{b}$, are reconstructed as two jets from the $b$
quarks, one leptonically decaying tau, and one hadronically decaying tau.
At $95\%$ C.L., we set a limit of $m(\VLQ) > 317 \gevcc$ for a model
with Yang-Mills type couplings~\cite{lq3YM}.
Figure~\ref{fig:lq3_and_lq2}(left) shows the cross section limits as a
function of $\VLQ$ mass for two models.

\subsection{$3^{\rm rd}$ Generation Scalar Leptoquarks}

Complementary to the search mentioned above, leptoquarks may also appear
as spin=0 objects.  A search at D\O\ assumes $3^{\rm rd}$ generation scalar
leptoquarks ($\SLQthree$) are pair produced and decay primarily via
$\SLQthree\to\nu b$, but also allows for $\SLQthree\to\tau t$~\cite{lq3d0}.
The considered decay products, $\nu b \overline{\nu} \overline{b}$, yield a
signature of
$2$ $b$-jets and $\met$ from the neutrinos. Based on $310 \invpb$, the
$95\%$ C.L.\ limit is $m(\SLQthree) > 213 \gevcc$.  If $\SLQthree\to\tau t$
is explicitly forbidden, the limit is $m(\SLQthree) > 219 \gevcc$.
Figure~\ref{fig:lq3_and_lq2}(right) shows these results.

\subsection{$2^{\rm nd}$ Generation Scalar Leptoquarks}

\begin{figure}
\psfig{figure=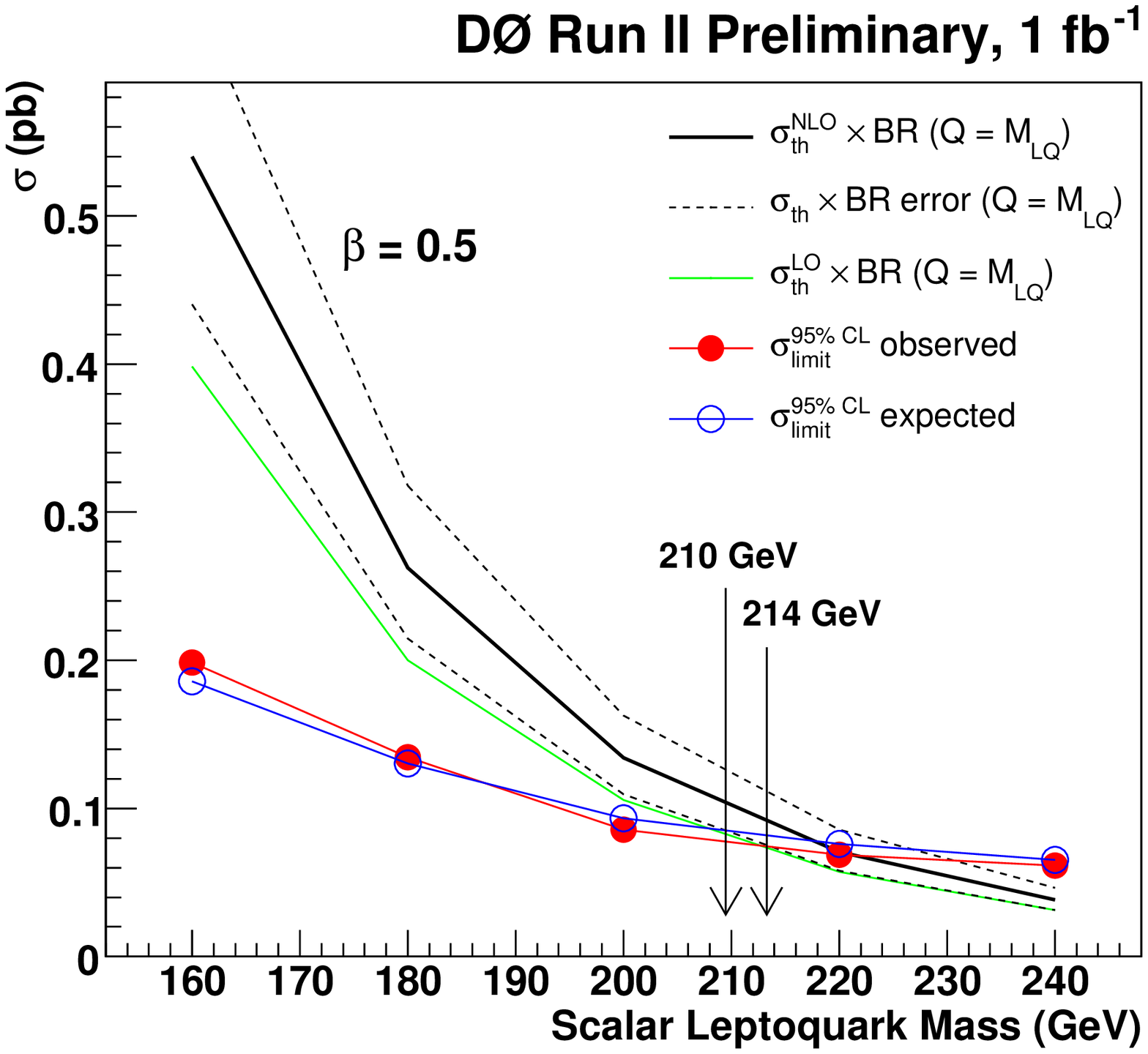,height=2.4in}
\psfig{figure=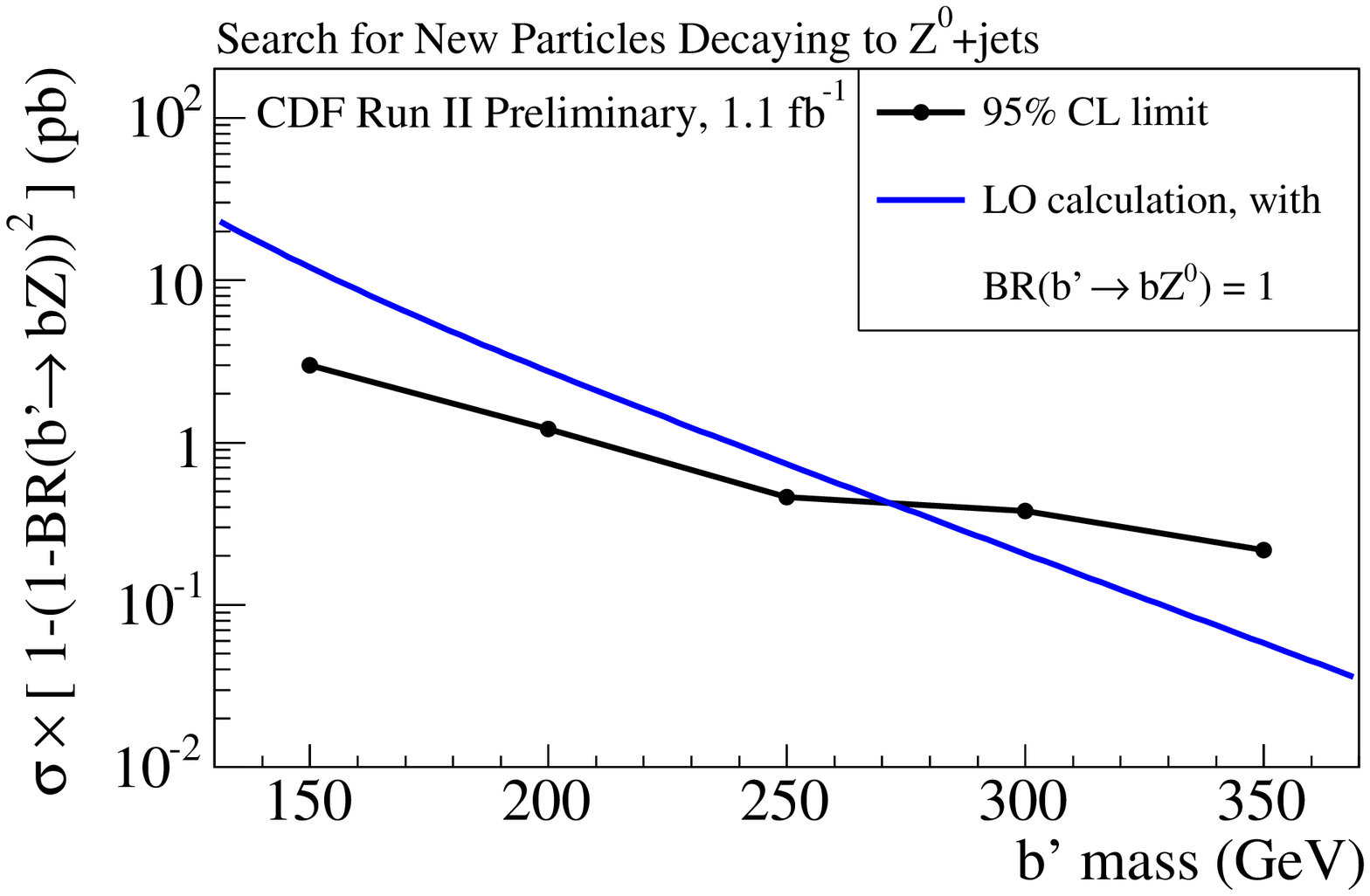,height=2.4in}
\caption{(left) Cross section limit for $2^{\rm nd}$ generation scalar
leptoquarks as a function of mass.  (right) Observed limit and theoretical
prediction for a $4^{\rm th}$ generation $b'$ model, in the search
for new particles that couple with $Z$ bosons.
\label{fig:slq2_and_bprime}}
\end{figure}

As with the other leptoquark searches, to avoid producing flavor changing
neutral currents, it is common to assume that generation number is conserved
in the decays.  Therefore a search for $2^{\rm nd}$ generation scalar
leptoquarks ($\SLQtwo$) at D\O\ assumes the decays contain muons or muon
neutrinos, and second generation quarks ($q = s$ or $c$)~\cite{lq2d0}.  We
consider the branching ratio $\beta \equiv {\cal B}(\SLQtwo\to\mu q) = 0.5$,
so that pair production and decay leads to a final state of $\mu q \nu q$.
The signature is one muon, $2$ jets, and $\met$.  In $1 \invfb$,
$6.1 \pm 1.1$ events are expected from SM backgrounds, while $6$ events
are observed.  The $95\%$ C.L.\ limit is $m(\SLQtwo) > 214 \gevcc$,
as shown in figure~\ref{fig:slq2_and_bprime}(left).

\section{Search for $b'$}

A search for $b'$ $4^{\rm th}$ generation quarks is carried out at CDF, as a
specific case to quantify the acceptance, in a general search for new
particles that couple to $Z$ bosons and jets~\cite{bprime}.  The production
and decay chain considered in the analysis is
$q\overline{q}\to g\to b'\overline{b}'\to bZ\overline{b}Z$, with one
$Z$ decaying through $Z\to\ell\ell$ and the other decaying through
$Z\to q\overline{q}$.  The leptonically decaying $Z$ is reconstructed in
the $Z\to ee$ and $Z\to \mu\mu$ channels, with lepton $\pt > 20 \gevc$,
and we require $\geq 3$ jets with $\Et > 30 \gev$.
The leading background, from SM $Z$+jet events, is estimated by using
events with $<3$ jets to predict the contribution for events with
$\geq 3$ jets.  Using $1.1 \invfb$, the $95\%$ C.L.\ limit is
$m(b') > 270 \gevcc$, as indicated by the intersection of the observed
and theoretical curves in figure~\ref{fig:slq2_and_bprime}.

\section{Large Extra Dimensions}

\begin{figure}
\begin{center}
\psfig{figure=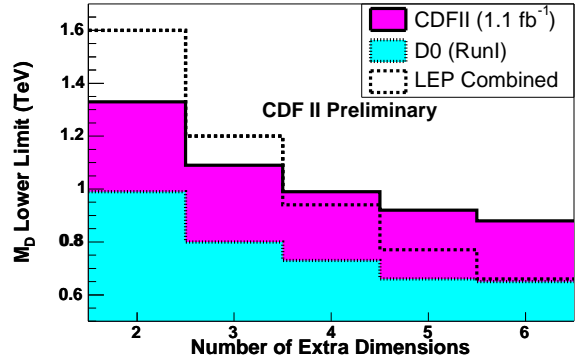,height=1.9in}
\end{center}
\caption{Lower limits on the effective Planck scale as a function of
the number of large extra dimensions.
\label{fig:led}}
\end{figure}

Large extra dimensions have been proposed~\cite{add} as a potential solution
to the hierarchy problem between the weak and gravitational forces.  In these
models, only gravitons ($G$) can propagate in the $n$ extra dimensions of
the $4+n$ dimensional bulk of space-time.  The Planck scale ($M_{\rm Planck}$)
is so large, and the strength of gravity so small, due to the large extent
($R$) of the extra spacial dimensions.  To address the hierarchy problem, the
new effective Planck scale ($M_D$) is related to the other quantities through
$M_{\rm Planck}^2 \sim R^n M_D^{2+n}$.  The gravitons can be directly
produced via $qq\to gG$, $qg\to qG$, or $gg\to gG$, in all cases yielding an
energetic jet from the quark or gluon, and $\met$ from the graviton.
The measurement at CDF, using $1.1 \invfb$, observes $779$ events while
expecting $819 \pm 17$ events from SM sources.  This is an
update, using the same selection but more data, of the previously
published results~\cite{led}.  These new results are used
to place lower limits on $M_D$ and upper limits on $R$.  The lower limits
on $M_D$ are given in figure~\ref{fig:led}, with corresponding upper limits
on $R$ of $R<0.27 \mm$, $R<3.1\times10^{-6} \mm$, $R<9.9\times10^{-9} \mm$,
$R<3.2\times10^{-10} \mm$, and $R<3.1\times10^{-11} \mm$ for $n=2,3,4,5$,
and $6$.

\section{Conclusions}
Various searches for new physics at D\O\ and CDF using photons and jets
are constraining specific models, while signature based searches are
looking for deviations from the SM.  Discoveries may be just around the
corner, as additional analyses and larger data sets are considered.

\section*{Acknowledgments}
We thank our colleagues with the CDF and D\O\ collaborations, the
Moriond conference organizers and participants, and the funding
agencies for making this work possible.  We also thank the European
Union Marie Curie Program for an accommodation grant.

\end{document}